# Universal unitary holonomies in photonics

Youlve Chen[1], Jinlong Xiang[1], Riyao Song[2,3], Yufeng Zhang[4], An He[1], Yunru Fan[2,3], Zhenyu Zhao[1], Yuchen Yin[1], Qiang Zhou[2,3], Guang-Can Guo[2,3], Yikai Su[1], Zhipei Sun[4†] and Xuhan Guo[1†]

[1]State Key Laboratory of Photonics and Communications, School of Information and Electronic Engineering, Shanghai Jiao Tong University; Shanghai, 200240, China.

[2]Institute of Fundamental and Frontier Sciences, University of Electronic Science and Technology of China, Chengdu 611731, China

[3]Center for Quantum Internet, Tianfu Jiangxi Laboratory, Chengdu 641419, China

[4]Department of Electronics and Nanoengineering, Aalto University, Espoo 02150, Finland.

*Corresponding author. Email: zhipei.sun@aalto.fi; guoxuhan@sjtu.edu.cn

## Abstract

Non-Abelian holonomy, a geometric aspect of quantum mechanics, underpins the geometric formulation of fundamental interactions and plays a vital role in holonomic quantum computation. Photonics is a fertile platform for exploring non-Abelian holonomies. However, universal unitary holonomies are still out of reach. The lack of control over scale phase degrees of freedom hinders the direct quantum simulation of unitary-governed phenomena—such as electron spin, weak isospin, and quantum chromodynamics—as well as the realization of generic geometric quantum gates for holonomic quantum computing. Here we employ artificial gauge fields (AGFs) to introduce complex-valued couplings, synthesizing arbitrary unitary holonomies and enabling generic geometric quantum gates in photonics. Higher-dimensional U($m$) transformations are experimentally synthesized based on Givens rotations. Our results pave the way for 'all-geometric-phase' approach for both quantum and classical computing in photonics.

## MAIN TEXT

## INTRODUCTION

Non-Abelian holonomy is a manifestation of the geometric structure of non-Abelian gauge theories, which underpin the description of weak and strong interactions between fundamental particles. In terms of applications, non-Abelian holonomy enables robust, fluctuation-resilient unitary operations and forms the foundation of holonomic quantum computation. The associated non-Abelian geometric phase is a generalization of the scalar Pancharatnam–Berry phase(*1, 2*), which arises from the parallel transport of a set of degenerate states around a closed loop in parameter space(*3*). As the salient ingredient of the 'all-geometric' approach(*4, 5*), non-Abelian holonomy has been utilized to implement universal quantum gates in various platforms, including cold atoms(*6-8*), superconducting circuits(*9-11*), trapped ions(*4, 12*), and nitrogen vacancy centers(*5, 13-15*). Recently, classical wave flux, such as light and sound that possess additional internal degrees of freedom, has become an emerging platform for exploring non-Abelian holonomy(*16*). Judicious manipulation of Hamiltonian with chiral symmetry in acoustic(*17, 18*) and photonics(*19-27*) can generate the matrix-valued non-Abelian geometric phase. However, existing photonic holonomy has so far been restricted to the special orthogonal group (SO(m))(*19, 21, 22, 24, 25, 27*) due to the intrinsic positive-valued couplings between dissipationless optical waveguides. This restriction hinders the direct quantum simulation of SU(2)-governed phenomena such as electron spin and weak isospin, as well as SU(3)-based theories including quantum chromodynamics. Crucially for applications, full unitary group (U(m)) holonomy is the key ingredient of the all-geometric' approach for holonomic computation in photonics, as it enables universal geometric quantum gates and can generate arbitrary linear transformations. In contrast, SO(m) holonomy is confined to real orthogonal transformations and thus limited to specific tasks.

Generic U(m) holonomy in photonics hinges on generating arbitrary complex-valued couplings—a capability available to electrons(*4, 5, 9, 11, 14, 15*) but profoundly challenging for photons. Artificial gauge fields (AGFs), implemented through periodic modulation within the framework of Floquet engineering to synthesize the effective electromagnetic fields, offer the solution. In photonics, AGFs are typically realized through periodically modulated waveguides, such as helically bent structures. These have been successfully employed to manipulate light guiding(*28, 29*), and engineer photonic topological bands such as Floquet topological insulators(*30-34*). More recently, AGFs have been applied to control anomalous coupling behaviors between integrated waveguides, such as dispersionless coupling(*35, 36*) and suppression of coupling (zero coupling)(*37, 38*).

In this work, we propose a generic framework that can design arbitrary U(m) non-Abelian holonomy in photonics: we realize complex-valued couplings based on artificial gauge fields by employing periodically bent waveguides, analogous to the complex hopping of electrons in the presence of an external magnetic field. This strategy offers novel insights into the manipulation of optical dynamics. The engineered complex couplings are further introduced into a four-waveguide system with chiral symmetry. By gradually tuning the bending amplitude and period, both the modulus and argument of the complex coupling can evolve adiabatically, enabling the construction of a U(2) holonomy. Any U(2) matrix can be decomposed into fundamental rotations generated by Pauli matrices $\sigma_z$, $\sigma_y$, and a global phase induced by the identity matrix *I*. Among these, the rotation generated by $\sigma_y$ has been previously realized using positive-valued couplings(*21, 22, 25*); the rotation generated by $\sigma_z$, and the global phase induced by *I*, which require complex-valued couplings, are realized by inducing Berry phases for two degenerate modes. We experimentally demonstrate these fundamental operations, and synthesize a generic U(2) transformation by combining different rotations. Using this capability, we implement geometric quantum gates. Non-Abelian characteristics of U(2) holonomy are also demonstrated. Furthermore, higher-dimensional U(m) matrices can be systematically synthesized through cascaded and parallel combinations of U(2) transformations, following the framework of Givens rotations(*39*). Especially, integrated photonics provides a scalable platform for high-dimensional holonomy. As a proof of concept, we also experimentally implement a U(4) transformation. Our work enables universal holonomic computation in photonics, opening new avenues for robust, all-geometric-phase-based classical and quantum computing in photonics.

## RESULTS

### Artificial gauge fields induce complex coupling

The coupling coefficient between two dissipationless straight waveguides is a positive number, originating from the mode overlap with the dielectric perturbation(*40*). Therefore, the Berry-Wilzeck-Zee (BWZ) connection is proportional to $\sigma_y$, which restricts the two-dimensional holonomy to SO(2) group ($e^{i\alpha\sigma_y}$) in photonics. The further realization of full U(2) group requires more varieties of rotations, such as those generated by $\sigma_z$, $\sigma_x$, and the global phase induced by the identity matrix *I*. Achieving such transformations necessitates the introduction of complex-valued coupling coefficients.

To address this problem, we employ periodic bent waveguides to introduce the AGFs, thereby enabling effective complex couplings. This effective coupling coefficient is reciprocal ($\kappa_{AB} = \kappa_{BA}^{\dagger}$), and the system is Hermitian. Taking advantage of the high-index contrast of integrated photonics, silicon waveguides can bear very sharp bending with low loss(*41*), which facilitates realizing the periodic driving of bendings to introduce AGFs(*35-38*). We begin our analysis with the complex coupling between two waveguides. The effective complex couplings between two identical periodic bent waveguides can be expressed as: $\kappa_{\text{bend}} = \frac{\kappa_{\text{straight}}}{P} \int_0^P e^{-i\beta g \frac{dx_0(z)}{dz}} dz = |\kappa_{\text{bend}}| e^{i\varphi}$ (*35, 42, 43*). Where the bending profile is denoted by $x_0(z)$ with a period $P$. The symbol $g$ denotes the center-to-center separation between two waveguides, $\kappa_{\text{straight}}$represents the coupling between two straight waveguides with the same center-to-center separation $g$, and $\beta$ is the propagation constant; the detailed derivation can be found in the Supplementary Text (ST1). It can be observed that $\kappa_{\text{bend}}$ can be a complex number (Eq. S8-S9), and the argument and modulus of $\kappa_{\text{bend}}$ can be modulated by the period, amplitude of bending profile $x_0(z)$, as derived in Eq. S9. We simulate different $x_0(z)$ and make a comparison with the coupling between straight waveguides. In two straight waveguides with positive-valued coupling, the phase difference between the two waveguides is exactly $\pi/2$ (Eq. S10 with $\varphi$=0). While for AGFs induced complex-valued coupling, the phase between two waveguides varies from 0~$\pi$ (Eq. S10 and Fig. S3B-D). The theoretical calculation and FDTD simulation fit well with each other (Fig. S3A).

We now turn to the implementation of complex coupling in a three-waveguide stimulated Raman adiabatic passage (STIRAP) system(*44-48*), whose zero mode plays a crucial role in constructing non-Abelian holonomy. As discussed in Supplementary Text (ST2), in the conventional STIRAP with positive real coupling coefficients, the phase difference in the two outer waveguides is exactly $\pi$ when the zero mode is excited (Fig. S5A). In contrast, for STIRAP systems incorporating complex-valued couplings, the phase difference in two outer waveguides is not $\pi$ (Fig. S5B-D), and is directly determined by the argument of the complex coupling. The simulation results are in agreement with the theoretical predictions.

**U(2) holonomy**

We now incorporate complex coupling into the realization of holonomy. The adiabatic evolution of complex-valued coupling can be precisely controlled by gradually varying the amplitude, period, and profile of bending. Both the modulus and argument of complex coupling can be adiabatically modulated to accomplish a holonomy. As Fig. 1A-B shows, the U($m$) holonomy device consists of periodic bent silicon waveguides on two layers. The inset shows a magnified view of a basic U(2) holonomy device, which consists of four waveguides. The four waveguides have an identical bending profile $x_0(z)$, as AGF-induced coupling necessitates that the surrounding waveguides (A, B, C) maintain the same bending geometry as the central waveguide X. Moreover, having a uniform bending profile ensures no relative dynamic phase accumulation or relative loss among the waveguides, which can be eliminated under gauge transformation. Thus, the coupling between waveguides B and X, C and X is complex-valued due to the induced AGFs. To be specific, the coupling between waveguides A and X is a positive real number since the displacement is zero ($g$=0) in the direction of AGF (Fig. S4 and discussed in ST1). Thus, the Hamiltonian can be written as:

$$H(z) = \begin{bmatrix} \beta & |\kappa_{AX}| & |\kappa_{BX}|e^{i\varphi_{BX}} & |\kappa_{CX}|e^{i\varphi_{CX}} \\ |\kappa_{AX}| & \beta & 0 & 0 \\ |\kappa_{BX}|e^{-i\varphi_{BX}} & 0 & \beta & 0 \\ |\kappa_{CX}|e^{-i\varphi_{CX}} & 0 & 0 & \beta \end{bmatrix} \quad (1)$$

Where, $|\kappa_{iX}|$, $\varphi_{iX}$ represent the modulus and argument of coupling, respectively. This Hamiltonian with chiral symmetry(*49*) supports two degenerate modes (Eq. S13). And the BWZ connection is also calculated in Eq. S15. Notably, the BWZ connection contains all three Pauli matrices ($\sigma_x$, $\sigma_y$, $\sigma_z$) as well as the identity matrix $I$, indicating that it can generate a U(2) holonomy. However, the expression is too complicated to directly map onto a holonomic path that realizes an arbitrary target matrix. Therefore, we aim to simplify the formulation and identify a generic framework that enables the systematic design of holonomic paths.

A generic U(2) matrix can be decomposed as four fundamental building blocks, fully containing four degrees of freedom for the U(2) group(*50*):

$$\mathrm{U}(2)=\exp(\mathrm{i}\alpha_1 I)\cdot\exp[\mathrm{i}\alpha_2\sigma_z]\cdot\exp(\mathrm{i}\alpha_3\sigma_y)\cdot\exp[\mathrm{i}\alpha_4\sigma_z] \qquad (2)$$

As illustrated in Fig. 1C-D, the U(2) rotation is visualized and decomposed on the Bloch sphere. Every element of the U(2) group can be expressed as the product of a U(1) phase factor ($\mathrm{e}^{\mathrm{i}\alpha_1 I}$) and an element of the SU(2) group. The SU(2) group, which represents three-dimensional rotations, can be further decomposed into successive rotations around the $z$-axis ($\mathrm{e}^{\mathrm{i}\alpha_2\sigma_z}$), $y$-axis ($\mathrm{e}^{\mathrm{i}\alpha_3\sigma_y}$), and $z$-axis ($\mathrm{e}^{\mathrm{i}\alpha_4\sigma_z}$)—commonly referred to as the $z$-$y$-$z$ Euler-angle decomposition(*51-55*). Thus, realizing these fundamental building blocks and cascading them sequentially can construct a generic U(2) holonomy. Among these rotations, $\mathrm{e}^{\mathrm{i}\alpha_3\sigma_y}$ that belongs to SO(2) group has been demonstrated in the reported scheme(*21, 22, 25*) with positive-valued coupling. Next, we will discuss the other rotations by introducing the Berry phase for dark modes through complex coupling.

We start with analyzing the Berry phase acquired in a three-waveguide STIRAP process involving complex-valued couplings. As Fig. 2A shows, the dark mode $|D_2\rangle$ experiences a three-waveguide STIRAP process, and $|D_1\rangle$ resides in an isolated waveguide. $|D_2\rangle$ can be expressed as $|D_2\rangle = -\mathrm{e}^{\mathrm{i}\varphi_{\mathrm{CX}}}\sin(\frac{\theta}{2})|\mathrm{A}\rangle + \cos(\frac{\theta}{2})|\mathrm{C}\rangle$. Where we define $\frac{\theta}{2} = \arctan(|\kappa_{\mathrm{CX}}|/|\kappa_{\mathrm{AX}}|)$. Thus, the Berry connection can be expressed as $A_{\varphi_{\mathrm{CX}}} = \mathrm{i}\langle D_2|\,\partial_{\varphi_{\mathrm{CX}}} D_2\rangle = -\sin^2(\frac{\theta}{2})$, $A_\theta = \mathrm{i}\langle D_2|\,\partial_\theta D_2\rangle = 0$, and there is a Berry phase $\oint -\sin^2(\frac{\theta}{2})\mathrm{d}\varphi_{\mathrm{CX}}$ acquired through the holonomy. It is evident that the corresponding Berry curvature can be analogous to the field of a magnetic monopole(*56*) with a charge of -1/2 located at the origin. Consequently, the Berry phase corresponds to negative half the solid angle $\alpha$ enclosed by the holonomic path on the Bloch sphere in Fig. 2A, and the star represents the start/end point of a holonomic path. An example of the simulated Berry phase, acquired through a holonomy in both Hilbert space and real space, is shown in Fig. S7 and discussed in ST2. In contrast, if the coupling is restricted to positive-valued, no geometric phase is accumulated over the two STIRAP processes, as the solid angle is zero (Fig. S6). Now we take the second degenerate dark mode $|D_1\rangle$ into consideration, this holonomy generates a basic unitary matrix of U(2) transformation for $|D_1\rangle$ and $|D_2\rangle$: $\begin{bmatrix} 1 & 0 \\ 0 & \mathrm{e}^{-\mathrm{i}\frac{\alpha}{2}} \end{bmatrix}$.

This scheme allows the implementation of multiple phase gates, such as the S and T gate. Fig. 2A also shows an example of the measured and predicted elements of this building blocks, and the detailed parameters, including the gap distance between waveguides, bending profiles of waveguides, as well as the evolution of complex-valued couplings are shown in Fig. S8A. The fidelity $F = 0.9427$, where, the fidelity is defined as $F = (1/N)\,|\mathrm{Tr}(U_\mathrm{t}^\dagger U_{\mathrm{exp}})|$(*57-60*) based on Frobenious inner product, Tr($A$) represents the trace of $A$, $N$ is the dimension of the unitary matrix, $U_\mathrm{t}$, $U_{\mathrm{exp}}$ are the theoretical and experimental unitary matrix, respectively.

If we also introduce the same Berry phase to $|D_1\rangle$, as illustrated in Fig. S14A, we cascade an identical STIRAP process for $|D_1\rangle$. The holonomic path of this STIRAP process for $|D_1\rangle$ in the Bloch sphere is also shown in Fig. S14A. In this way, both degenerate two modes acquire the same geometric phase. Consequently, a fundamental building block of the U(2) group: $\mathrm{e}^{-\mathrm{i}\frac{\alpha}{2}\boldsymbol{I}} = \begin{bmatrix} \mathrm{e}^{-\mathrm{i}\frac{\alpha}{2}} & 0 \\ 0 & \mathrm{e}^{-\mathrm{i}\frac{\alpha}{2}} \end{bmatrix}$ is generated. Fig. S14A shows an example of these building blocks, and the detailed parameters of the holonomy is shown in Fig. S8B. The fidelity between experimental results and theoretical prediction is 0.9288.

We now focus on the remaining building blocks of U(2) group in Eq. 2: $\mathrm{e}^{\mathrm{i}\alpha\sigma_z}$ and $\mathrm{e}^{\mathrm{i}\alpha\sigma_y}$. For the $\mathrm{e}^{\mathrm{i}\alpha\sigma_z}$ component, a straightforward approach is to introduce two Berry phases with opposite signs for $|D_1\rangle$ and $|D_2\rangle$, as the abovementioned method. However, in order to reduce the device length, we adopt another configuration. As illustrated in Fig. 2B, the structure is designed as $g_{\mathrm{BX}} = g_{\mathrm{CX}}$ and $\kappa_{\mathrm{BX}} = \kappa_{\mathrm{CX}}^{\dagger}$. Consequently, the BWZ connection in Eq. S15 can be significantly simplified, and the two degenerate modes $|D_1\rangle$ and $|D_2\rangle$ acquire geometric phases with identical absolute values but opposite signs (Eq. S17-S21). The geometric phase is the solid angle enclosed by the holonomic path on the Bloch sphere in Fig. 2B. As a result, a fundamental building block of U(2) transformation: $\mathrm{e}^{\mathrm{i}\alpha\sigma_z} = \begin{bmatrix} \mathrm{e}^{\mathrm{i}\alpha} & 0 \\ 0 & \mathrm{e}^{-\mathrm{i}\alpha} \end{bmatrix}$ is generated. An exemplified experimental and theoretical results of this rotation is shown Fig. 2B, the detailed parameters of the holonomy is illustrated in Fig. S8C, and the fidelity is 0.9830.

We now address the final building block, $\mathrm{e}^{\mathrm{i}\alpha\sigma_y} = \begin{bmatrix} \cos(\alpha) & \sin(\alpha) \\ -\sin(\alpha) & \cos(\alpha) \end{bmatrix}$, which belongs to SO(2) group and has been demonstrated in reported schemes(*21, 22, 25*). In these configurations, three positive-valued couplings are isomorphic to a 2-sphere, and the corresponding geometric phase is determined by the solid angle $\alpha$ enclosed by the holonomic path on that sphere. The experimental and theoretical results are shown in Fig. S14B, and the fidelity is 0.9832.

The above experiments, performed with classical light, measure the output intensity and phase profiles to characterize the implemented unitary operations. Notably, we measure the detailed phase of individual matrix elements through the interference method (some phase values are labeled with "N/A" since their theoretical intensity is zero, rendering the phase devoid of physical meaning). For example, to measure the geometric phase of a beam undergoing the holonomy, we measure the phase difference between light propagating through the holonomy and that through an isolated reference waveguide (Fig. S9C-D). The reference waveguide shares the same bending profile as the holonomy waveguides and therefore accumulates the same dynamical phase. By measuring the relative phase between the two paths, we can extract the geometric phase of matrix elements eliminating the dynamical phase contribution. Besides, we also measure the relative phase between different matrix elements (Fig. S9A-B). More information about the interference measurement methodology can be found in ST4.

To ensure the completeness of the U(2) group representation, we refer to the Euler parametrization in Eq. 2. If $\alpha_1$, $\alpha_2$, and $\alpha_4$ span the range from -π to π, and $\alpha_3$ covers the interval from -π/2 to π/2 , this is sufficient to realize full coverage of U(2) group(*51*). The range of $\alpha_3$ has been demonstrated in reported works. Negative values of $\alpha_1$, $\alpha_2$, and $\alpha_4$ can be implemented by simply exchanging $P_1$ and $P_2$. Furthermore, larger effective rotation angles can be achieved by cascading multiple holonomic devices in sequence.

Furthermore, we exemplify a U(2) transformation that incorporates multiple types of rotations into a device, as shown in Fig. 2C. At step 1, $|D_1\rangle$ resides in the isolated waveguide B, while $|D_2\rangle$ experiences a three-waveguide STIRAP process involving waveguides A, X, and C. $\kappa_{\mathrm{CX}}$ starts with a complex value, and adiabatically evolves to a positive value. Therefore, $|D_2\rangle$ acquires a Berry phase. Step 2 involves all positive-valued coupling, and the BWZ connection is proportional to $\sigma_y$, contributing to an SO(2) rotation between $|D_1\rangle$ and $|D_2\rangle$. At step 3, $|D_2\rangle$ resides in the isolated waveguide C, while $|D_1\rangle$ experiences a three-waveguide STIRAP process composed of waveguides A, X, and B. $\kappa_{\mathrm{BX}}$ starts with a positive value, and adiabatically evolves to a complex value. Thus, $|D_1\rangle$ acquires a Berry phase. The theoretically predicted and experimentally measured matrix elements are also shown in Fig. 2C-D.

## SU(2) geometric quantum gates

Based on the aforementioned complete U(2) holonomic transformation, we implement single-qubit geometric quantum gates using heralded single photons. As Fig. 3A shows, entangled photon pairs are generated in a periodically poled $LiNbO_3$ (PPLN) waveguide. The signal photon is routed to the chip while the idler photon is directed to the superconducting nanowire single-photon detectors (SNSPD) for heralding. More details of the entangled photon pairs source can be found in Ref. (*61*). The quantum gates are characterized by the process matrices $\chi$, and the evolution from the initial state $\rho_i$ and to the final state $\rho_f$ can be described by a quantum dynamic map: $\rho_f = \sum_{m,n} \chi_{mn} E_m \, \rho_i E_n^{\dagger}$(*5, 9, 15, 62*). For single-qubit gates, the full set of orthogonal bases is chosen as the Pauli basis $[I, X, Y, Z]$=$[I, \sigma_x, \sigma_y, \sigma_z]$(*5, 62*). The process matrix $\chi$ is reconstructed by quantum process tomography (QPT)(*62*). QPT requires preparing different input states and performing projective measurements on the output states, corresponding to the state preparation and state analysis(*63*) modules in Fig. 3A. Therefore, Mach-Zehnder interferometer is employed, and the thermal phase shifter (heater) is characterized(*63*). Two holonomic gates are demonstrated: $\mathrm{R_y}(-\frac{\pi}{2}) = \frac{1}{\sqrt{2}}\begin{bmatrix} 1 & 1 \\ -1 & 1 \end{bmatrix}$ and the phase gate $\mathrm{R_z}(-\frac{\pi}{2}) = \begin{bmatrix} 1 & 0 \\ 0 & -\mathrm{i} \end{bmatrix}$ ($S^{\dagger}$ gate), and their theoretically predicted and experimentally measured process matrices $\chi$ are shown in Fig. 3B-C. The process fidelity is defined as: $F = \mathrm{Tr}(\chi_{\mathrm{exp}}\chi_{\mathrm{theo}})$(*5*), and the fidelities are 0.9182 and 0.9274, respectively.

## Non-Abelian characteristic of U(2) holonomy

To further illustrate the Abelian nature of SO(2) and the non-Abelian character of U(2), we perform cascaded holonomic operations. Figure 4A presents the composition of two successive SO(2) rotations, which commute with each other, demonstrating the Abelian property. In contrast, Fig. 4B demonstrates the composition of a U(2) and an SO(2) rotation, which exhibits non-commutative behavior, confirming the non-Abelian nature of the U(2) holonomy.

## High-dimensional U($m$) holonomy

Any target U($m$) matrix can be decomposed into a cascade and parallel combination of U(2) matrices using Givens rotations(*39*). As Fig. 5A shows, we exemplify a U(3) matrix through a mesh comprised of three U(2) matrices. The mathematical model, experimental results, and theoretical prediction are also shown in Fig. 5A, and the fidelity is 0.9832.

Figure 5B shows a U(4) matrix through a mesh comprised of six U(2) matrices. The mathematical model, experimental results, and theoretical prediction are also shown in Fig. 5B, and the fidelity is 0.9524. Here, we only measure the intensity of U($m$) elements since it is very challenging to conduct $m^2$ times of interference to acquire the phase of all elements.

## DISCUSSION

The manipulation of dynamic and geometric phases plays a fundamental role in information processing within electronic and photonic systems. Compared with the dynamic phase, the 'all-geometric' approach brings the intrinsic robustness to quantum computation against the environmental noise and decoherence. The successful realization of SU(m) or U(m) groups has enabled fault-resilient universal holonomic quantum gates across various platforms. Moreover, scalability remains a long-term goal for achieving efficient multi-qubit manipulation. While geometric-phase-based (holonomic) quantum computing has been well established in various systems, it has remained elusive in photonics, where most approaches rely on dynamic phase control — such as in Mach-Zehnder interferometer meshes(*64*), micro-ring resonator arrays(*65*), diffractive metasurfaces(*66, 67*), and intensity modulation(*57, 68*) schemes. In contrast, an all-geometric approach via holonomy in photonics provides broader bandwidth for high-dimensional matrices(*25*) and greater fabrication tolerance(*25*), as it relies only on the path in parameter space and benefits from adiabatic evolution.

In conclusion, we resolve the key issue of non-Abelian holonomy in photonics—construction of generic unitary group to fulfill the demand of universal quantum gates or arbitrary classical linear computing tasks in photonics. Besides, the excellent scalability, and mass-production integrated photonic platform enables the large-scale computing networks. This work demonstrates the feasibility of implementing geometric-phase-based computing in photonic systems, opening a promising route toward inherently fault-tolerant photonic quantum computing and novel high-performance classical optical processors.

## MATERIALS AND METHODS

### Device fabrication

The fabrication process of device is compatible with CMOS technology. The devices were fabricated on a 500-μm-thick silicon substrate wafer with a 3-μm-thick buried oxide layer. For the first layer, a 100-nm-thick amorphous silicon thin film was deposited by plasma-enhanced chemical vapor deposition (PECVD, Oxford Instruments). The PECVD gas included $SiH_4$, and Ar. Then photoresist (AR-P 6200.09) was spun at a speed of 4000 r/min to have a thickness of 200 nm on the chip, and baked at a temperature of 180°C for 2 mins. The designed patterns were defined by electronic beam lithography (EBL, Vistec EBPG-5200), and the developing uses methyl isobutyl ketone (MIBK, developing for 75 secs) and isopropanol (IPA, developing for 60 secs). Then it was fully etched through inductively coupled plasma deep reactive ion etching system for silicon (ICP Deep RIE, SPTS Technologies). After that, the chip was coated with silica as inter-layer dielectrics using inductively coupled plasma chemical vapor deposition (ICPCVD). For the second layer, similar processes were repeated as the first layer, such as depositing 100-nm-thick amorphous silicon film, defining patterns through EBL, fully ICP Deep RIE, etc. Finally, a 1-μm-thick silica cladding was deposited through PECVD. The fabrication process is summarized in Fig. S1.

For the fabrication of phase shifter (including titanium heater, golden electrical wires and electrodes), first we spun the photoresist Poly(methyl methacrylate) (PMMA) at a speed of 4000 r/min to have a thickness of 900 nm on the chip, and baked at a temperature of 180°C for 2 mins. The designed heater patterns were defined by EBL. After developing, 200-nm-thick titanium thin film is deposited through electron beam evaporation (Denton Vacuum). After metal deposition, a standard lift-off process was conducted by immersing the chip in acetone for several hours, followed by gentle sonication to remove the residual resist and unwanted metal patterns. Then similar process is conducted for golden conductor path and pad, including spun the photoresist, EBL, developing, depositing golden thin film, and lift-off.

### Measurement setup

#### Classical experiments

The light is emitted from a wideband source (Wuhan Leicheng Photonics Technology Co., Ltd.). Then it propagates through a polarization controller and is coupled in/out to the integrated photonic chip. Edge couplers are utilized to interface the light between single-mode fiber and silicon waveguide. The light output is monitored by a power meter (JW3216, Shanghai Joinwit) and the spectra are analyzed through OSA (AQ6370C, YOKOGAWA). In addition, we use a precision power supply (Keithley 2400) as the direct current (DC) power source.

#### Quantum experiments

Photon pairs are generated through second-harmonic generation (SHG) and spontaneous parametric down-conversion (SPDC) in a piece of pigtailed fiber periodically poled $LiNbO_3$ (PPLN) waveguide. More details on the generation of entangled photon pairs can be found in Ref.(*61*). One photon is injected into the device. The other single photon is injected into a piece of single-mode fiber. Then the signal and idler photons are detected by two superconducting nanowire single-photon detectors (SNSPDs, P-CS-6, PHOTEC). A time-to-digital converter (TDC, ID900, ID Quantique) is used to record the counting rates and the coincidence events between the signal and idler photons(*61*). These are summarized in Fig. 3A.

**Acknowledgments**

We would like to thank the Center for Advanced Electronic Materials and Devices (AEMD) of Shanghai Jiao Tong University (SJTU) and Tianjin H-chip Technology for the support in device fabrication. We also thank Prof. Ian H. White (previous in University of Cambridge & University of Bath) for fruitful discussion.

**Funding:**

National Key R&D Program of China grant 2023YFB2804702

Natural Science Foundation of China (NSFC) grant 62550072, 62341508, 62505175, 625B2113, 62405185

Shanghai Science and Technology Innovation Action Plan (Grant 25LN3201000, 25JD1405500, 24JD1401500)

Shanghai Municipal Science and Technology Major Project

**Author contributions:**

Conceptualization: XHG, YLC

Methodology: XHG, YLC, RYS, YRF, QZ

Investigation: XHG, YLC, RYS, YRF

Visualization: XHG, YLC, JLX, AH, YCY

Funding acquisition: XHG

Project administration: XHG

Supervision: XHG, ZPS, QZ, GCG, YKS

Writing – original draft: XHG, YLC, JLX, AH

Writing – review & editing: XHG, ZPS, YLC, YKS

**Competing interests:** Authors declare that they have no competing interests

**Data and materials availability:** The datasets generated during the current study are available in the Figshare repository

## Figures

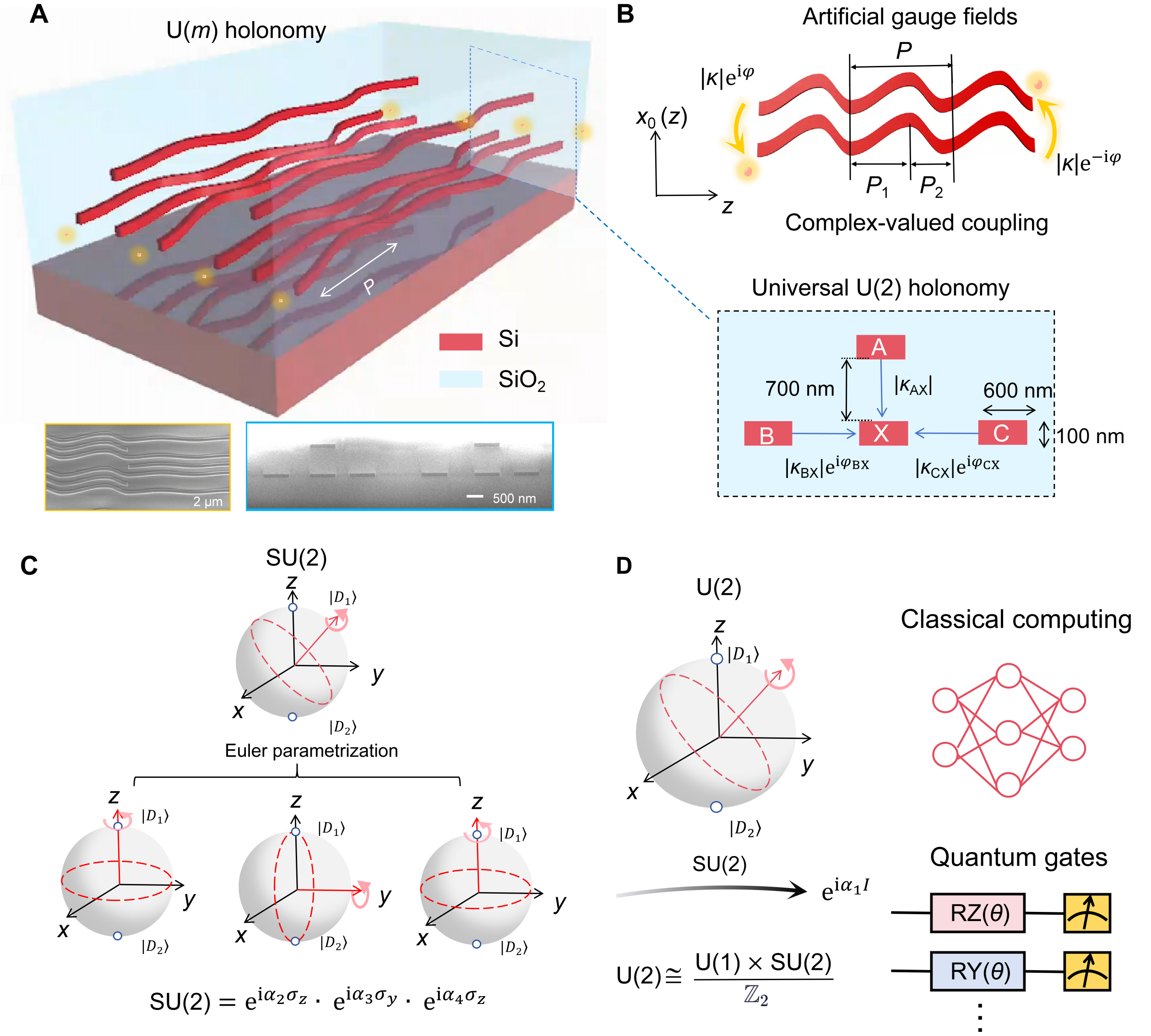

**Fig. 1. Generic U($m$) holonomy. (A)-(B)** Structure of U($m$) holonomy device enabled by AGFs, with an inset magnification illustrating the structure of a basic U(2) holonomy device. **(A)** Schematic of a U($m$) holonomy device. The insets are scanning electron microscope (SEM) figures and the focused ion beam (FIB) image in top-view and cross-sectional view, respectively. **(B)** The U(2) holonomy device is comprised of four periodic bent silicon waveguides A, B, C, and X on two layers. All four waveguides have an identical bending profile $x_0(z)$. Thus, the coupling between waveguides B and X, C and X are complex-valued due to the induced AGF. The adiabatic evolution of complex-valued coupling can be precisely controlled by slowly varying the amplitude, period as well as profile of bending. **(C)-(D)** Illustration and decomposition of U(2) rotation in the Bloch sphere. **(C)** SU(2) group can be further decomposed as the rotation around the $z$-axis, $y$-axis, and $z$-axis. **(D)** Every element of U(2) group can be decomposed as the product of $e^{i\alpha I}$ and an element of SU(2) group.

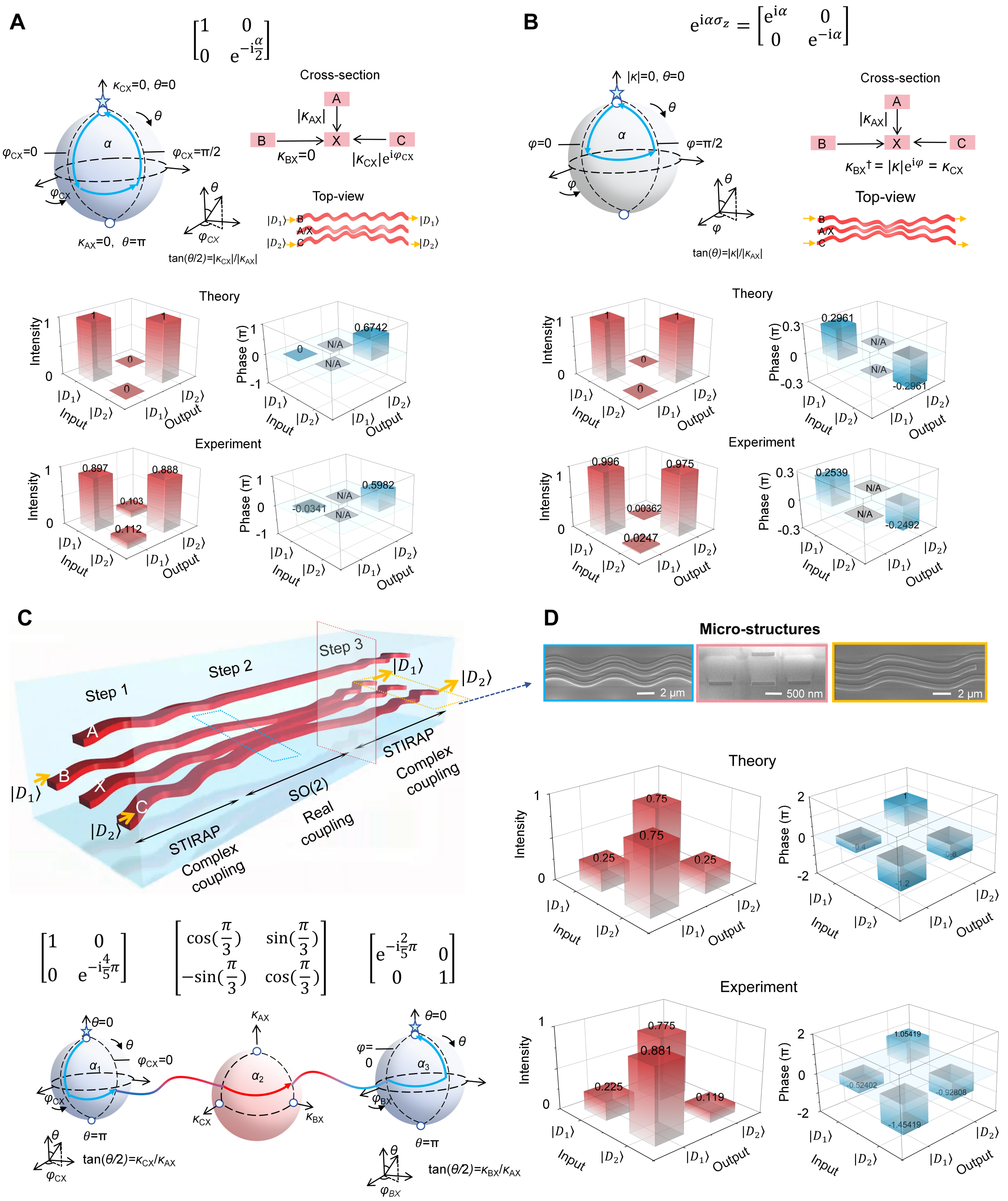

**Fig. 2. Experimental results, theoretical prediction, holonomic paths in Hilbert space as well as structures in real space for different building blocks of U(2) transformation. (A)** $\begin{bmatrix}1 & 0\\ 0 & e^{-i\frac{\alpha}{2}}\end{bmatrix}$, $|D_1\rangle$ resides in an isolated waveguide, $|D_2\rangle$ acquires the Berry phase $-\frac{\alpha}{2}$ after experiencing two STIRAPs with complex-valued coupling. The detailed parameters can be found in Fig. S8A. **(B)** $e^{i\alpha\sigma_z} = \begin{bmatrix}e^{i\alpha} & 0\\ 0 & e^{-i\alpha}\end{bmatrix}$, the structure is designed as $g_{BX} = g_{CX}$ and $\kappa_{BX} = \kappa_{CX}^{\dagger}$. Two degenerate modes $|D_1\rangle$ and $|D_2\rangle$ acquire Berry phases with identical absolute values but opposite signs. The detailed parameters can be found in Fig. S8C. **(C)-(D)** A U(2) holonomy that involved different kinds of rotations. **(C)** The structure in real space and the holonomic path in Hilbert space. **(D)**

Experimental results and theoretical prediction. The insets are SEM figures and the FIB image in top-view and cross-sectional view, respectively.

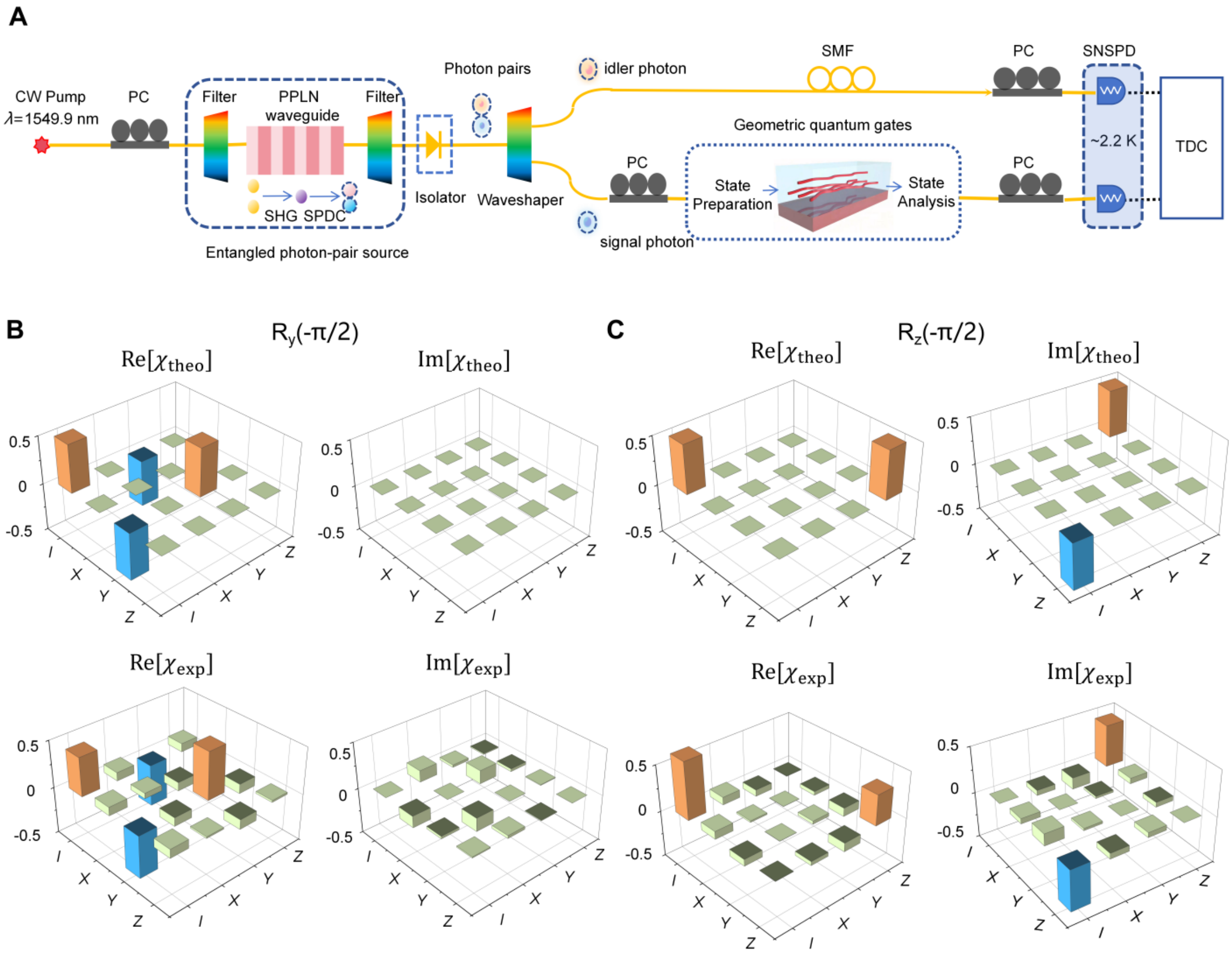

**Fig. 3. Experimental results of geometric quantum gates.** **(A)** Experimental setups for the measurement of holonomic gates. SHG, second-harmonic generation. SPDC, spontaneous parametric down-conversion. PPLN, periodically poled $LiNbO_3$. PC, polarization controller. TDC, time-to-digital converter. SNSPD, superconducting nanowire single-photon detectors. SMP, single-mode fiber. CW, continuous wave. More details of the entangled photon pairs source can be found in Ref. 60. **(B)-(C)** Theoretically predicted and experimentally measured process matrices $\chi$ of two single-qubit geometric quantum gates. **(B)** $R_y(-\pi/2)$ gate. **(C)** $R_z(-\pi/2)$ gate ($S^\dagger$ gate).

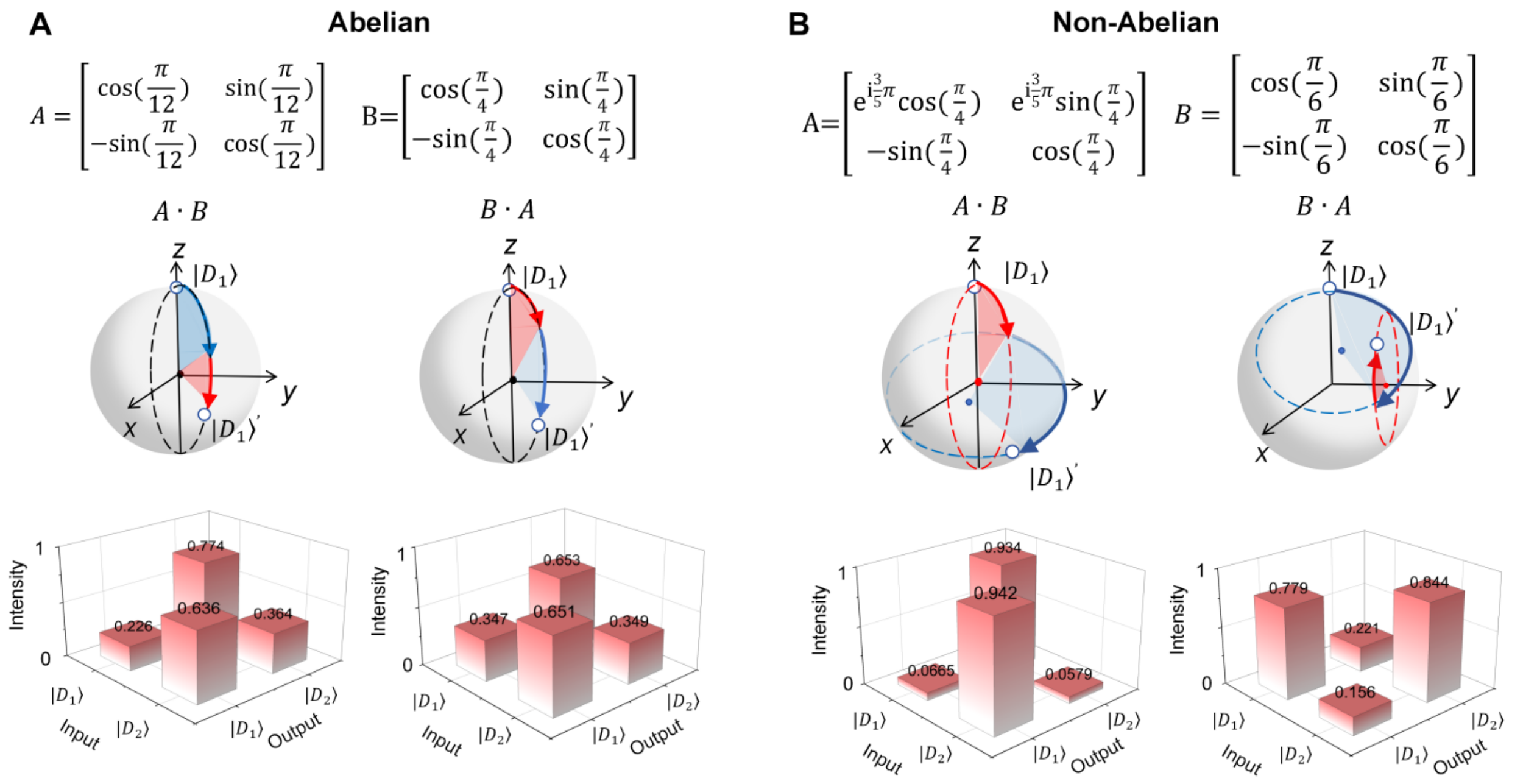

**Fig. 4. Demonstration of non-Abelian characteristics. (A)-(B)** Demonstration of the non-Abelian characteristic. **(A)** Cascading of two SO(2) rotations ($e^{i\alpha\sigma_y}$), which are Abelian. **(B)** Cascading of a U(2) rotation and a SO(2) rotation, they are non-Abelian.

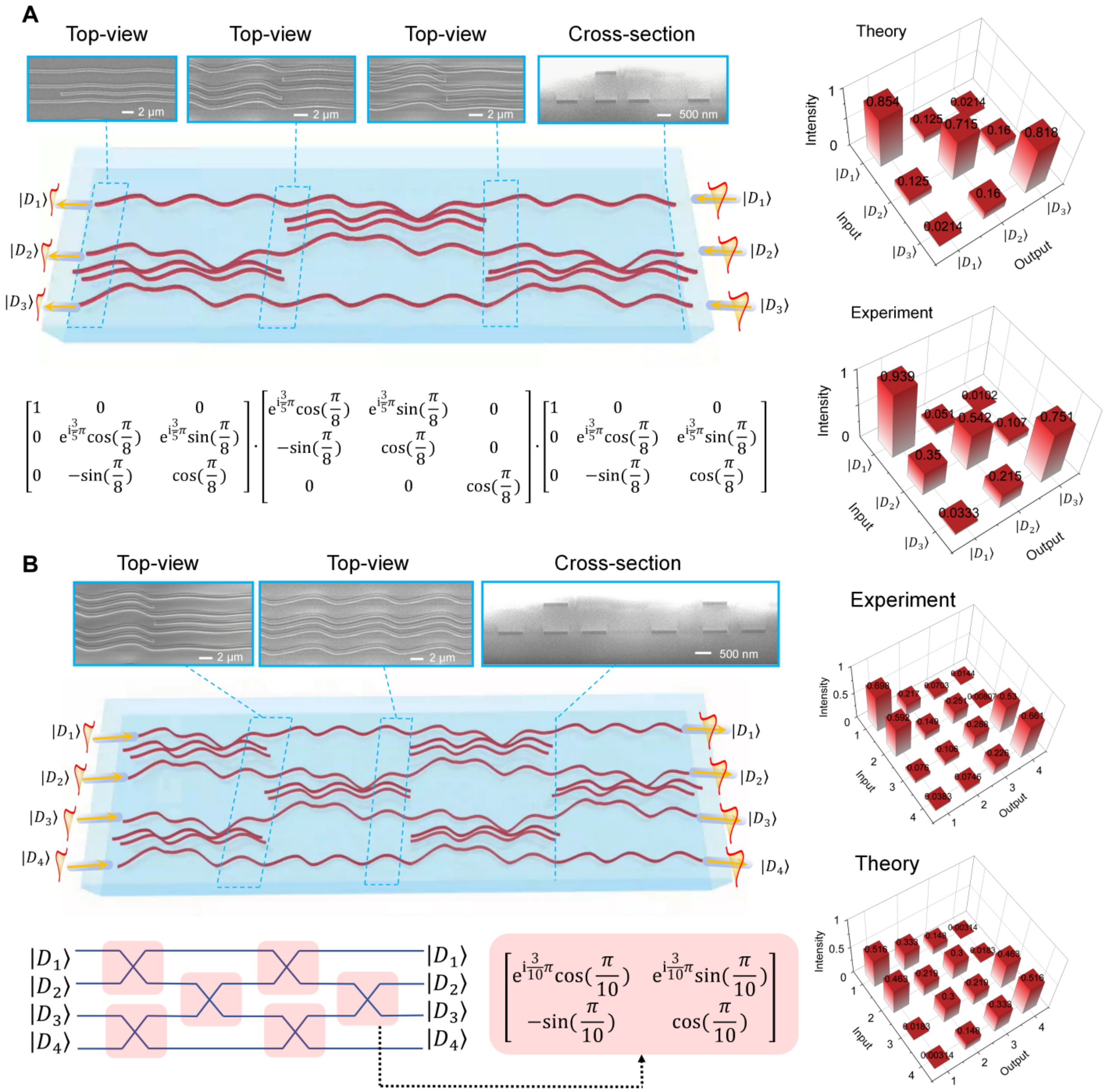

**Fig. 5. Experimental results of U(3) and U(4) holonomy. (A)** U(3) holonomy. The experimentally measured and theoretically predicted intensity, structure, as well as the mathematical model are shown. The insets are SEM figures and the FIB image in top-view and cross-sectional view, respectively. **(B)** U(4) holonomy. The experimentally measured and theoretically predicted intensity, structure, as well as the mathematical model are shown. The insets are SEM figures and the FIB image in the top-view and cross-sectional view, respectively.

**Supplementary Materials**

Supplementary Text

Figs. S1 to S14

References (*1–24*)